\documentclass[pra,preprint,preprintnumbers,amsmath,amssymb]{revtex4}
\usepackage{textcomp}
\usepackage{amssymb}
\usepackage{amsmath}
\usepackage{amsthm}

\usepackage{graphicx}
\usepackage{txfonts}

\begin{document}

\preprint{}
\title{Generation of perfect vortex and vector beams based on Pancharatnam-Berry phase elements}
\author{Yachao Liu}
\author{Yougang Ke}
\author{Junxiao Zhou}
\author{Yuanyuan Liu}
\author{Hailu Luo}\email{hailuluo@hnu.edu.cn}
\author{Shuangchun Wen}
\author{Dianyuan Fan}
\affiliation{Laboratory for Spin Photonics, School of Physics and
Electronics, Hunan University, Changsha 410082, China}

\date{\today}
\maketitle

\textbf{Perfect vortex beams are the orbital angular momentum
(OAM)-carrying beams with fixed annular intensities, which provide a
better source of OAM than traditional Laguerre-Gaussian beams.
However, ordinary schemes to obtain the perfect vortex beams are
usually bulky and unstable. We demonstrate here a novel generation
scheme by designing planar Pancharatnam-Berry (PB) phase elements to
replace all the elements required. Different from the conventional
approaches based on reflective or refractive elements, PB phase
elements can dramatically reduce the occupying volume of system.
Moreover, the PB phase element scheme is easily developed to produce
the perfect vector beams. Therefore, our scheme may provide
prominent vortex and vector sources for integrated optical
communication and micromanipulation systems.}

Orbital angular momentum (OAM) has been identified as a useful
degree of freedom for enhancing the information carrying capacity of
photon because of its unlimited dimensions~\cite{Allen1992}, and
also has been found to have broad applications in optical
manipulation and metrology due to the singularities in phase and
intensity~\cite{Grier2003}. Laguerre-Gaussian (LG) beam is the most
widely studied optical carrier of OAM as its helical phase
($exp(il\varphi)$, $l$ is the topological charge, and $\varphi$ is
the azimuthal angle) contributes to a finite OAM per photon
($l\hbar$, $\hbar$ is the reduced Plank's constant
$h/2\pi$)~\cite{Allen1992}. However, general LG modes are
characterized with an annular intensity varying with the change of
topological charge, which inevitably confines the co-propagation of
multiple OAM modes in communication systems. Thus, a reliable source
providing OAM-carrying beams with fixed size of intensity would be
meaningful. Moreover, a OAM source with large topological charge and
small size of ring-pattern intensity is generally expected for
trapping and manipulating the nano-particles.

Perfect vortex (PV) beams provide a solution to this
problem~\cite{Ostrovsky2013}, which keep the size of intensity
pattern no matter what the topological charge is. Theoretically, PV
beam is the Fourier transformation of a Bessel Gaussian (BG) beam
~\cite{Vaity2015}. Plenty of schemes have been proposed to obtain
the PV beams including using spatial light
modulator~\cite{Ostrovsky2013,Vaity2015,Garcia-Garcia2014,Arrizon2015,Banerji2016},
axicon~\cite{Chen2013,Jabir2016}, interferometer~\cite{Li2016}, and
micro-mirror devices~\cite{Zhang2016}. Optical manipulation based on
the tightly focused field of PV beams has also been extensively
studied~\cite{Chen2013,Paez-Lopez2016}. The utility of particle
trapping and high efficient OAM transfer promise a broad application
scope of PV beams. Furthermore, the non-diffraction property of PV
beams scattering field shows the potential use in imaging and
cryptography~\cite{Reddy2016}. However, most of these methods are
dependent on cumbrous reflective devices or bulky refractive devices
to produce PV beams, which undoubtedly limits the application of PV
beams in miniaturized optical systems and in fiber communications.

In this work, we designed planar Pancharatnam-Berry (PB) phase
elements to generate PV beams. The evolution of polarizations in
optical field will introduce an additional phase item in wave
function, which is referred as the PB
phase~\cite{Pancharatnam1956,Berry1984}. PB phase optical elements
are dramatically developed in recent years due to their exceptional
property of phase engineering, which provide a convenient approach
to construct highly-functional planar
devices~\cite{Lin2014,Liu2017}. Arbitrary phase distributions can be
pursued and the phase responses of different incident circular
polarizations are constantly contrary in a PB phase element.
Therefore, PB phase elements including optical
lens~\cite{Hasman2003}, vortex beam
generators~\cite{Bomzon2002,Biener2002,Marrucci2006}, hologram
devices~\cite{Huang2013}, mode transformer~\cite{He2015}, and OAM
multiplexer~\cite{Mehmood2016} have been realized up to now. By
designing and assembling novel planar PB phase elements, a simplest
scheme to produce PV beams is presented in this work. Moreover, our
scheme is developed to generate the perfect vector beams, which
possesses also the fixed annular intensity but is endowed with
azimuthally varying polarizations.
\\

\begin{figure}
\centering
\includegraphics[width=12cm]{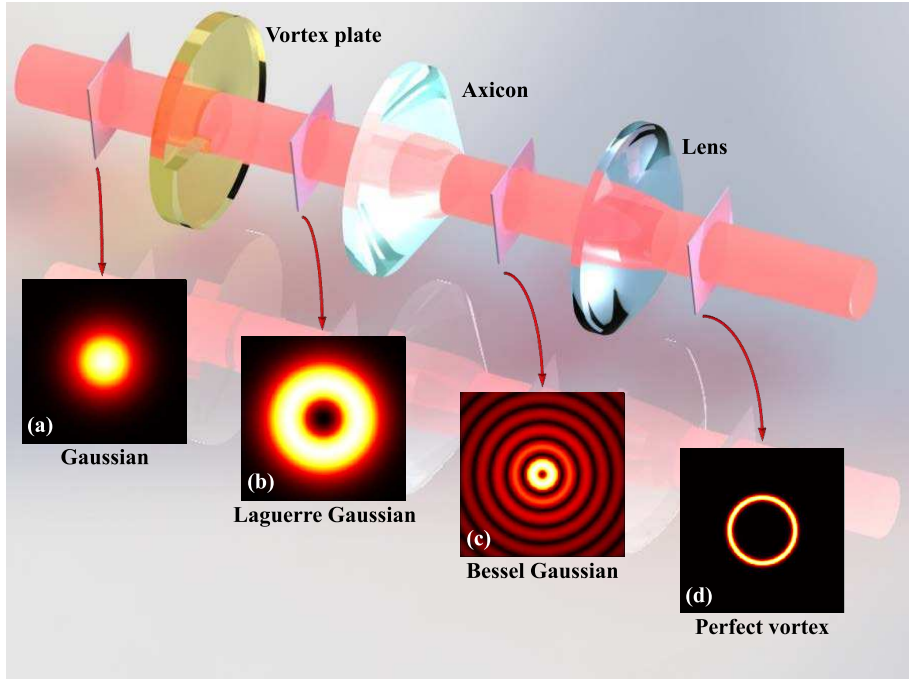}
\caption{\label{Fig1} The conventional optical system applied to
generate PV beams, which is based on the bulky refractive devices.
Combination of vortex plate and axicon convert a general Gaussian
beam to BG mode. Then, the lens implements Fourier transformation of
BG beam to get the PV beams. Insets show the intensity patterns in
each step of this system.}
\end{figure}

\noindent\textbf{RESULTS}\\
\noindent\textbf{Theoretical analysis}. Perfect vortex beams are
firstly considered as an ideal model of vortex beams with constant
intensity distribution which doesn't depend on its topological
charge~\cite{Ostrovsky2013}. Dirac delta function was applied to
describe these beams as the following form:
\begin{equation}
E(\rho,\varphi)=\delta(\rho-\rho_{0})\exp(i l \varphi)\label{delta},
\end{equation}
where $(\rho,\varphi)$ are the polar coordinates in beam cross
section, $l$ is the topological charge, and $\rho_{0}$ is the radius
of annular bright intensity. However, in general cases, this model
is not accessible in experiment. Thus, an annual pattern with small
width $\Delta\rho$ is developed to approximate the above model,
which can be described as
\begin{equation}
E(\rho,\varphi)=\exp
\left[-\frac{(\rho-\rho_{0})^{2}}{\Delta\rho^{2}}\right] \exp(i l
\varphi)\label{exp}.
\end{equation}

\begin{figure}
\centering
\includegraphics[width=12cm]{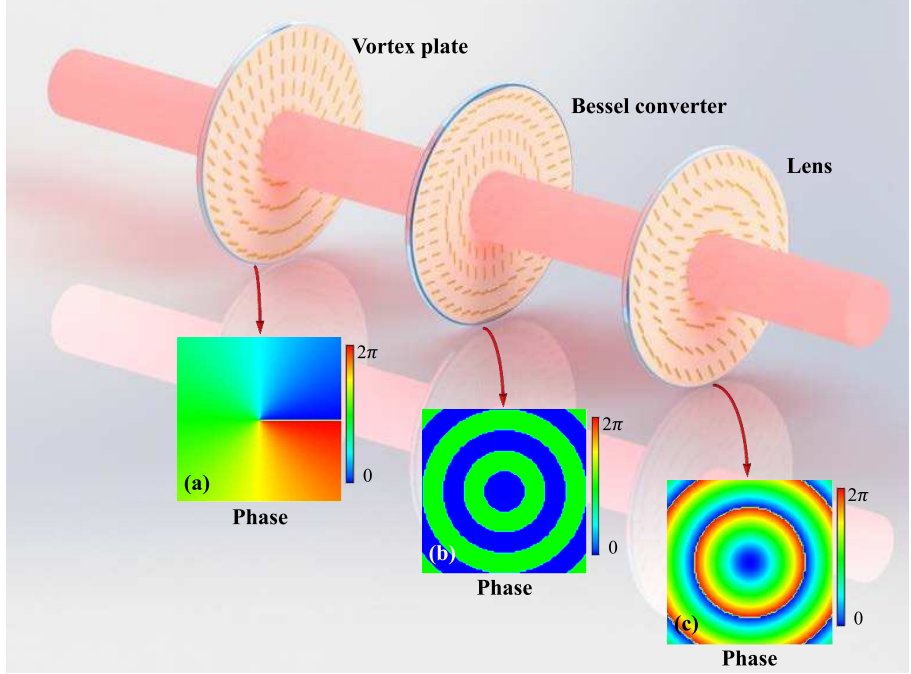}
\caption{\label{Fig2} The PB phase element system used to produce PV
beams. Planar vortex plate and Bessel converter are combined to
generate a higher-order BG mode. Following lens transfers this mode
to correspond PV beam. Comparing with the system exhibited in
Fig.~\ref{Fig1}, PB phase elements will dramatically reduce the
occupying volume of the entire system. Insets show the phase
distributions of each element respectively.}
\end{figure}

Mathematically, the approximate model of PV beams can be deduced
from the Fourier transformation of an ideal Bessel beam
function~\cite{Vaity2015}, which can be expressed as
\begin{equation}
E_{B}(r,z)=J_{l}(k_{r} \rho)\exp(i l \varphi+ik_{z}z)\label{Bessel},
\end{equation}
where $J_{l}$ is the first kind of $l$-th order Bessel function,
$k=\sqrt{{k_{r}}^{2}+{k_{z}}^{2}} = {2 \pi}/{\lambda}$,
$r=(\rho,\varphi)$, and $(k_{r}, k_{z})$ are the radial and
longitudinal wave vectors respectively. However, this mathematical
model is also an idealization of PV beams. For most part of
practical situations, the accessible Bessel model is actually a BG
beam:
\begin{equation}
E_{BG}(\rho,\varphi)=J_{l}(k_{r} \rho)\exp(i l \varphi)\exp
\left(-\frac{\rho^{2}}{{\omega_{g}}^{2}} \right)\label{BesselG},
\end{equation}
where $\omega_{g}$ is the beam waist to restrict the total field.
Many approaches have been exploited to generate such BG beams.
Axicon is the most widely applied refractive
element~\cite{Arlt2000}, which could convert a general Gaussian beam
or LG beams to the zeroth or higher-order BG modes.

According to the Fourier transformation theory, the transformed
field of a BG beam can be derived as
\begin{equation}
E(\gamma,\vartheta)=i^{l-1}\frac{\omega_{g}}{\omega_{0}}\exp(il\vartheta)\exp
\left(-\frac{\gamma^{2}+\gamma_{r}^{2}}{\omega_{0}^{2}} \right)I_{l}
\left(\frac{2\gamma_{r}\gamma}{\omega_{0}^{2}} \right)\label{PV},
\end{equation}
where $\omega_{0}=2f/k\omega_{g}$ is the Gaussian beam waist at
focal plane, $f$ is the focal length, $I_{l}$ is the $l$-th order
modified Bessel function of the first kind. It is clear that a
modified Bessel function and a Gaussian function are conjoined to
shape the amplitude, which contribute to the maximum value at
$\gamma=\gamma_{r}$ and the sharp reduction of amplitude beyond the
range $\omega_{0}$. Thus, a vortex beam with fixed annular intensity
(radius equals $\gamma_{r}$ and width equals $2\omega_{0}$) is
constructed, which is namely the PV beam. The value of annular
radius $\gamma_{r}$ depends on the radial wave vector, which can be
modulated by the parameter of axicon
($\gamma_{r}=k_{r}f/k=f\sin[(n-1)\beta]$, where $n$ is the
refractive index and $\beta$ is the base angle of axicon).

Figure~\ref{Fig1} demonstrates a set of refractive apparatus for
producing PV beams. A vortex plate constructed with increasing
thickness along azimuthal direction, which gives beam the phase
variation from $0$ to multiple $2\pi$, will transform a general
Gaussian beam (Inset (a) in Fig.~\ref{Fig1}) to higher-order LG
beam~\cite{Beijersbergen1994}, as the example shown in inset (b) of
Fig.~\ref{Fig1}. Then, an axicon possessing conical surface is
applied to obtain the corresponding BG mode (Inset (c) in
Fig.~\ref{Fig1}). Finally, a plano-convex lens is used to implement
the Fourier transformation and the PV beam is produced as shown in
inset (d) of Fig.~\ref{Fig1}.

\begin{figure}[!hb]
\centering
\includegraphics[width=12cm]{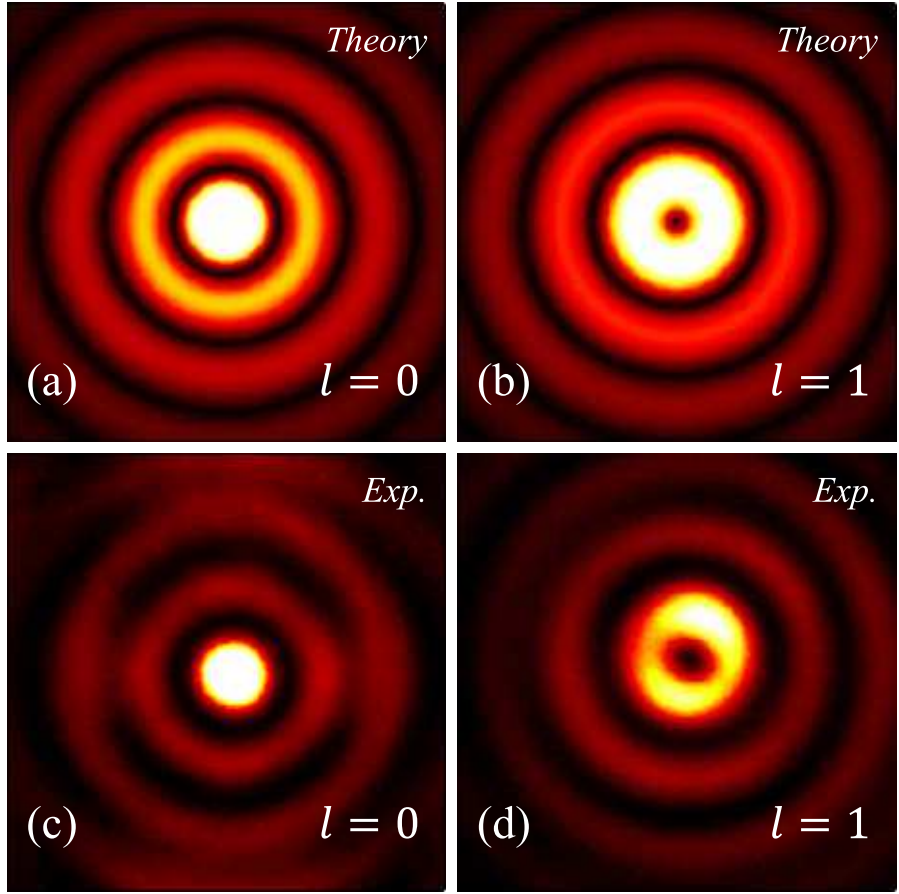}
\caption{\label{Fig3} Intensities of zeroth (left column) and first
order (right column) BG modes obtained by using planar PB phase
elements. First line and second line show the theoretical and
experimental distributions respectively.}
\end{figure}

To replace all the elements involved in the conventional refractive
system, we should firstly find out the phase profiles imposed by
these elements. Moreover, comparing with using axicon in producing
the BG beam, there is a simpler scheme by designing a planar device
to introduce the derived phase from Bessel function (as shown in
inset (b) of Fig.~\ref{Fig2})~\cite{Vaity2015}. Thus, the phase
distributions corresponding to vortex plate $\Phi_{V}$, Bessel
converter $\Phi_{B}$, and lens $\Phi_{L}$ can be defined as
\begin{equation}
\Phi_{V}(x,y)=m\cdot \arctan \left(\frac{y}{x}\right)\label{vp},
\end{equation}
\begin{equation}
\Phi_{B}(x,y)=\arg[E_{B}(r,z)]\label{bp},
\end{equation}
\begin{equation}
\Phi_{L}(x,y)=\frac{2\pi}{\lambda}\left[\sqrt{(x^{2}+y^{2})+f^{2}}-f\right]\label{axiconp},
\end{equation}
where $(x,y)$ is the Cartesian coordinates in beam cross section,
$m$ is the topological charge of desired mode, $E_{B}(r,z)$ is the
function of Bessel beam presented in Eq.~\ref{Bessel} when $l=0$,
and $f$ is the focal length of lens. Examples of phase distributions
($m=1$) are demonstrated in insets of Fig.~\ref{Fig2}.

Pancharatnam-Berry phase elements are widely studied in recent years
as their responses can be locally adjusted by rotating the
orientation of building blocks. When a PB phase element is
constructed based on the local anisotropic response of blocks, the
geometric phase obtained in a circular polarization incidence can be
concluded as $\Phi=\pm2\phi$~\cite{Hasman2003}, where the sign
depends on the helicity of the circular polarization, and $\phi$
refers to orientation local optical axis. Thus, arbitrary phase
devices can be contrived according to this relationship. The axis
orientations of vortex plate, Bessel converter, and lens are deduced
as
\begin{equation}
\phi_{V}(x,y)=\frac{m}{2}\cdot\arctan\left(\frac{y}{x}\right)\label{vpo},
\end{equation}
\begin{equation}
\phi_{B}(x,y)=\frac{1}{2}\arg[E_{B}(r,z)]\label{bpo},
\end{equation}
\begin{equation}
\phi_{L}(x,y)=\frac{\pi}{\lambda}\left[\sqrt{(x^{2}+y^{2})+f^{2}}-f\right]\label{axiconpo}.
\end{equation}
Figure~\ref{Fig2} shows the schematics of these PB phase elements
and their assembly to produce PV beams. Short lines in each plate
suggest the orientation of local optical axes. Comparing with the
conventional case, the PB phase system is simplified as the great
reduction of thickness of each element.\\

\begin{figure}[!hb]
\centering
\includegraphics[width=12cm]{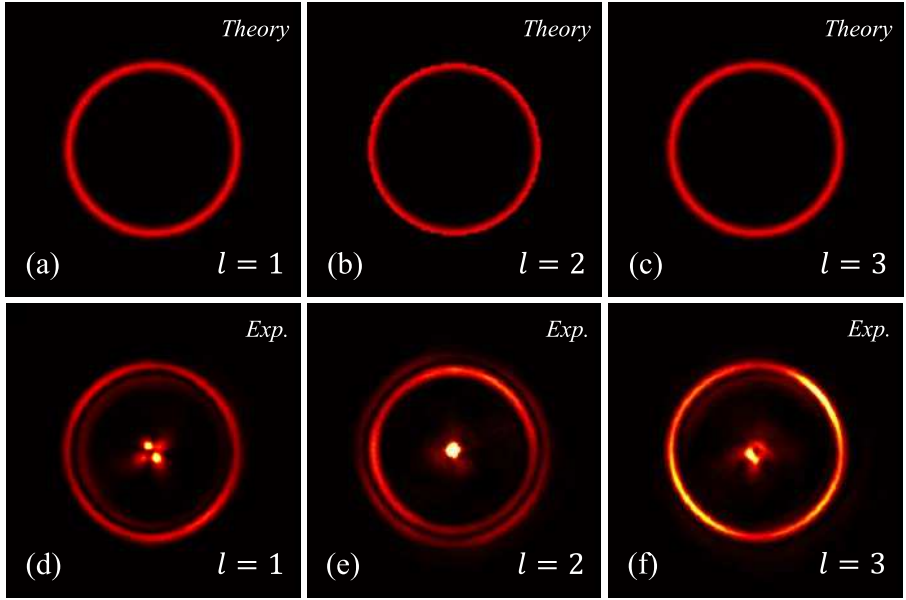}
\caption{\label{Fig4} The PV beams generated in experiment. Three
different topological charges, $l=1$, $2$, and $3$ are chosen. First
line shows the theoretical patterns, and second line shows the
experimental results respectively. The same as theoretical
predictions, radiuses of observed annular intensities are
independent on the topological charges.}
\end{figure}

\begin{figure}[!h]
\centering
\includegraphics[width=11cm]{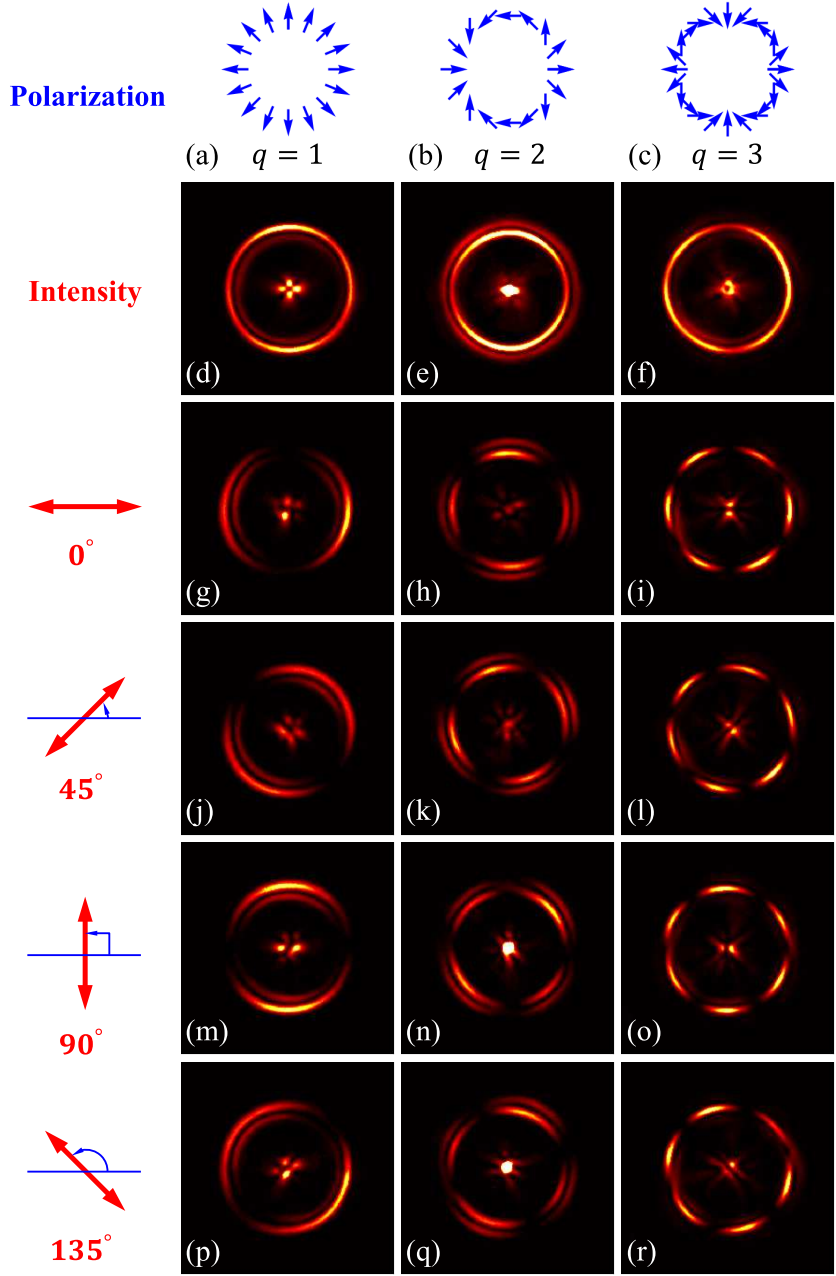}
\caption{\label{Fig5} The perfect vector beams generated in our
experiment. Three different polarization orders $q=1$, $2$, and $3$
are displayed from left to right. First line shows the schematic
polarization distributions (blue arrow diagrams), of which the
experimental results are presented in each column respectively.
Second row is the observed intensities which show the similar
annular patterns as corresponding PV beams. Following rows are the
polarizer examined intensities to verify the polarization
distributions. Different axis orientations of polarizer $0^{\circ}$,
$45^{\circ}$, $90^{\circ}$, and $135^{\circ}$ (red arrows in the
first column) are selected in experiment. Well agreements suggest
that the obtained perfect beams are CVB, namely the PV beams here.}
\end{figure}

\noindent\textbf{Experimental results.} By using the PB phase
elements designed above, BG beams and PV beams were produced in our
experiments. The combination of PB phase vortex plate and Bessel
converter produced high quality BG beams in experiments as presented
in Fig.~\ref{Fig3}. Theoretical intensities are obtain according to
Eq.~(\ref{BesselG}) where $\omega_{g}$ and $k_{r}$ are well selected
(as shown in the upper row of Fig.~\ref{Fig3}). Experiment results
of two different orders $l=0~and~1$ are demonstrated in the lower
row of Fig.~\ref{Fig3}): the first one was produced by the single PB
phase element constructed according to Eq.~(\ref{bpo}), of which the
optical axis distribution are exhibited as the second plate in
Fig.~\ref{Fig2}; the second one was produced by the combination of
vortex plate (according to Eq.~(\ref{vpo})) and the zeroth order BG
plate. Incident light was set to be circular polarization (either
right- or left-circular polarizations) by a combination of polarizer
and quarter-waveplate.

Then, the Fourier transformation was implemented by the PB phase
lens (according to Eq.~(\ref{axiconpo}), where $f$ are chosen as
$200mm$) to get PV beams. Annular intensities were observed on the
focal plane as shown in Fig.~\ref{Fig4}. Different results of
topological charges $l=1$, $2$, and $3$ are demonstrated. The upper
row shows the intensities calculated according to Eq.~(\ref{PV}) and
the lower row is the corresponding experimental results. The
radiuses of recorded annular intensities are unvaried when changing
the topological charges. Defects located at the center of patterns
are inevitable because a portion of diffused light was focused by
the exit lens.
\\

\noindent\textbf{DISCUSSION}\\
The advantage on the PB phase system over traditional system in
obtaining PV beams is not only the reduction of thickness of the
whole system but also the promising prospect in integration
applications. All these three elements can be once fabricated in a
single glass thus to provide a PV source in integration photonics
and plasmonics. Moreover, as the contrary phase responses of
orthogonal circular polarization states can be introduced at the
same time, our PB phase system can be directly adapted to produce
the perfect vector beams.

Cylindrical vector beam (CVB) is the axially symmetrical solution of
vectorial electromagnetic field Maxwell's equations~\cite{Zhan2009},
which have shown great potentials in particle
trapping~\cite{Zhan2004}, plasmon excitation~\cite{Zhan2006},
microscopy~\cite{Abouraddy2006}, high resolution
imaging~\cite{Novotny2001,Mino2016}, high capacity information
coding~\cite{Li2011}, and laser processing~\cite{Drevinskas2016}.
Generally, the intensity of CVB is endowed with a LG profile.
However, vectorial fields corresponding to other intensity profiles
may provide the additional exotic properties. For examples, vector
BG and Hermite-Gaussian have been studied recently to show their
distinctiveness~\cite{Pfeiffer2014,Chen2016,Fu2016}. PV beams
provide excellent resistance to the variation of topological charge
and the perturbations in propagation, thus, the CVB with PV profile
will be fervently desired in applications.

Cylindrical vector beam has been proved to be the superposition of
two orthogonal circularly polarized vortex beams with opposite
chirality~\cite{Milione2011,Liu2014}. By changing the incidence to
linear polarization, which is the equal superposition of orthogonal
circular polarizations, the combination of  PB phase vortex plate
and Bessel converter will produce a vector BG field. However, the PB
phase lens will terminate this adaption because contrary radial
phase gradients will be obtained by the opposite incidences. Thus, a
general plano-convex lens is applied to replace the PB phase lens to
get the perfect vector beams. Experimental results are displayed in
Fig.~\ref{Fig5}. Three different polarization orders $q=1$, $2$, $3$
are selected to examine the validity of our scheme. Polarizer
analysis shows that the correct polarizations as theoretical
predications are generated.

In conclusion, we designed a series of Pancharatnam-Berry phase
elements to produce the perfect vortex and vector beams. The
structures of Pancharatnam-Berry phase elements are derived and
improved from the conventional refractive perfect vortex generating
system. Moreover, by combining the advantages of PB phase elements
and refractive elements, we developed the scheme to generate perfect
vector beams. Well agreements are presented in our experiments. The
observed intensities are independent on the topological charges and
polarization orders. Both of perfect vortex and vector beams are the
candidate sources for large-capacity communication, high resolution
imaging, and many other potential applications. Our scheme may
provide the versatile solutions for
integrated optics and fiber optics.\\

\noindent\textbf{METHODS}\\
\noindent\textbf{Sample preparation}. Our PB phase elements are
fabricated using femtosecond laser writing in fused silicon glass
boards. With certain level intense femtosecond laser irradiation,
sub-wavelength period grating oriented perpendicular to the
polarization of laser will be formed spontaneously in
glass~\cite{Beresna2011}. Thus, the artificial birefringence is
introduced to mould the polarization of light. As known, the
principle optical axes are parallel or orthogonal to the
subwavelength grating, thus the orientation of local optical axes
can be artificially engineered by controlling the polarization of
writing laser. Additionally, other than structures wrote pixel by
pixel in general cases, these self-assembled structures can be
written continuously, which means higher transform efficiency can be
accessed. To get the PB phase samples, we firstly found out all the
elements required to generate PV beams. Then, as all the elements
are phase manipulating devices, their phase distributions are
mathematically described. After that, according to the relationship
between the orientation of local optical axes and the induced PB
phase, the distribution of local optical axes can also be well
described. Finally, all the elements are fabricated according to the
derivation. Our samples are constructed for the operating wavelength
$632.8nm$, efficient diameter $\geq6mm$ (Altechna R\&D).

\noindent\textbf{Experimental measurements}. The schematic of
experiment setup is shown in Fig.~\ref{Fig2}. All the optical
elements are cascaded to produce the PV beams. In experiments, a
He-Ne laser operating in wavelength $632.8nm$ (17 mW, Thorlabs
HNL210L-EC) served as the optical source. The combination of linear
polarizer and quarter-waveplate was used to prepare the polarization
state of incidence. A CCD (charge-coupled device, Coherent
LaserCam HR) camera was applied to record all the intensities.\\

\setlength{\parindent}{0pt} \textbf{Acknowledgements}\\This research
was supported by the National Natural Science Foundation of China (Grants No. 11274106 and No. 11474089).\\

\textbf{Author contributions}\\Y.L. and H.L. conceived the idea and
designed the experiment. The theoretical work and experiment were
performed by Y.L.. Y.K., J.Z., and Y.-Y.L. H.L., S.W., and D.F. supervised all aspects of the
project. All the authors contributed to the writing of the manuscript.\\

\textbf{Additional information}\\
Supplementary information accompanies this paper at
http://www.nature.com/scientificreports

Correspondence and requests for materials should be addressed to
H.L.\\ \\
Competing financial interests: The authors declare no
competing financial interests.

\end{document}